\documentclass[fleqn, usenatbib]{mnras}

\usepackage{newtxtext,newtxmath}	
\usepackage[T1]{fontenc}
\usepackage{ae,aecompl}

\usepackage{graphicx}	% Including figure files
\usepackage{amsmath}	% Advanced maths commands
\usepackage{amssymb}	% Extra maths symbols
\usepackage{csquotes}
\usepackage{textcmds}
\usepackage{natbib}
\newcommand{\kms}{km s$^{-1}$}

\title[Periodic OH masers ]{Periodic variability of the mainline hydroxyl masers in G9.62+0.20E}

\author[S. Goedhart et al.]{
S. Goedhart$^{1,2}$\thanks{E-mail: sharmila@ska.ac.za},
R. van Rooyen$^{1,2}$,
D.J. van der Walt$^{2}$, 
J.P. Maswanganye$^{2}$,
\newauthor
A. Sanna$^{3}$,
G.C. MacLeod$^{4,5}$
 and S.P. van den Heever$^{4}$
\\
$^{1}$South African Radio Astronomy Observatory, 2 Fir Street, Black River Park, Observatory, 7925, South Africa
\\
$^{2}$Center for Space Research, North-West University, Potchefstroom campus, Private Bag X6001, Potchefstroom, 2520, South Africa \\
$^3$  Max-Planck-Institut f{\"u}r Radioastonomie, Auf dem H{\"u}gel 69, D-53121 Bonn, Germany \\
$^4$ Hartebeesthoek Radio Astronomy Observatory, PO Box 443, Krugersdorp, 1740, South Africa\\
$^{5}$The University of Western Ontario, 1151 Richmond Street. London, ON
N6A 3K7, Canada\\
}

\date{Accepted 2019 March 7. Received 2019 February 27; in original form 2018 December 20}

\pubyear{2018}

\begin{document}
\label{firstpage}
\pagerange{\pageref{firstpage}--\pageref{lastpage}}
\maketitle

\begin{abstract}
We present the results of a monitoring campaign using the KAT-7 and HartRAO 26m
telescopes, of hydroxyl, methanol and water vapour masers associated with the high-mass
star forming region G9.62+0.20E. Periodic flaring of the main line hydroxyl masers were
found, similar to that seen in the 6.7 and 12.2 GHz methanol masers. The 1667 MHz flares are
characterized by a rapid decrease in flux density which is coincident with the start of
the 12.2 GHz methanol maser flare. The decrease in the OH maser flux density is followed
by a slow increase till a maximum is reached after which the maser decays to its pre-flare
level. A possible interpretation of the rapid decrease in the maser flux density is
presented. Considering the projected separation between the periodic methanol and OH
masers, we conclude that the periodic 12.2 methanol masing region is located about 1600 AU deeper
into the molecular envelope compared to the location of the periodic OH masers. A single
water maser flare was also detected which seems not to be associated with the same event
that gives rise to the periodic methanol and OH maser flares.

\end{abstract}


%%%%%%%%%%%%%%%%%%%%%%%%%%%%%%%%%%%%%%%%%%%%%%%%%%

%%%%%%%%%%%%%%%%% BODY OF PAPER %%%%%%%%%%%%%%%%%%

\section{Introduction}

To date there are at least 20 known periodic methanol masers \citep{Goedhart2009,
  Araya2010, Szymczak2011a, Goedhart2014, Fujisawa2014, Szymczak2015a, Maswanganye2015, Szymczak2016, Maswanganye2016, sugiyama2017} including one source which shows quasi-periodic variations in both
methanol and formaldehyde \citep{Araya2010}.  The formaldehyde and methanol variations
were simultaneous and showed very close correspondence to each other, which the authors
speculated was due to modulation of the infrared radiation field by periodic accretion in
a binary system.  \citet{Green2012} found a possible indication of periodic variability in
the hydroxyl masers in G12.89+0.49, where the hydroxyl masers may have undergone a
periodic drop in intensity. However, their results were not conclusive since the time
series was undersampled.  \citet{Szymczak2016} found anti-correlated variations
in the water and methanol masers in G107.298+5.639 even though there are good spatial and
velocity overlap between the two masers. Methanol and hydroxyl masers are thought to share a
similar (radiative) pump mechanism \citep{Cragg2002} and are often found in close spatial
proximity, while water masers are collisionally pumped and require high density
environments.  Thus, correlated variability may be expected between hydroxyl and methanol
masers.  However, since the two molecules have different energy level structures, they may
not necessarily respond in exactly the same way to changes in the pumping radiation
field. The fact that several of the periodic methanol maser sources also have associated
hydroxyl masers allows us to investigate to what extent the hydroxyl masers also show
periodic flaring and therefore to possibly determine whether the flaring is due to changes
in the maser amplification or in the background seed photon flux.

A number of known periodic methanol masers were monitored on a weekly basis at 1665 and
1667 MHz as part of the seven-element Karoo Array Telescope (KAT-7) science verification
programme. This paper focuses on G9.62+0.20E, which showed clear evidence of variability in
the hydroxyl masers during the first year of monitoring. It was subsequently monitored on
a daily basis during the expected June 2014 and February 2015 methanol maser flares while
simultanously monitoring the 6.7 and 12.2 GHz methanol and 22.2 GHz water masers using the
Hartebeesthoek Radio Astronomy Observatory (HartRAO) 26m telescope.

G9.62+0.20E is a high-mass star forming region with a hypercompact \ion{H}{II} region
harbouring at least one massive star in an early evolutionary phase \citep{Garay1993}.  It
has a number of masers projected against the \ion{H}{II} region - Class II methanol
masers, water masers and hydroxyl masers \citep{Sanna2015} and is located at a distance
of 5.2 kpc from the Sun \citep{Sanna2009}.  It was the first methanol maser source
discovered to show periodic variations \citep{Goedhart2003}, with a best-fit period of
243.3 days \citep{Goedhart2014}.  Simultaneous flares in methanol have been observed at
6.7-, 12.2- and 107 GHz \citep{VanderWalt2009}. One explanation of the methanol maser
flare profiles is by the variation of the free-free continuum flux in the background
\ion{H}{II} region due to variable ionizing radiation associated with shocked winds from a
colliding wind binary system with a non-zero orbital eccentricity \citep{VanderWalt2011}.
However, it has been argued by \citet{Parfenov2014a} that the same effect can be produced
by variations of the dust temperature in an accretion disk around a forming binary system,
while \citet{Inayoshi2013} suggest that pulsational instabilities of massive protostars
could arise during rapid accretion. \citet{Singh2012} investigated the possibility of
bipolar outflows in young binary systems as a possible explanation for some of the
periodic methanol masers. See \citet{vanderwalt2016} for a further discussion.

The OH monitoring of G9.62+0.20E revealed complex behaviour across multiple OH maser
features, with some features showing a pronounced drop in power at the same time that the
methanol masers started to flare. Not all of the hydroxyl or methanol maser features
flare, and since the relative positions of all the maser species and the background
\ion{H}{II} region is known \citep{Sanna2015}, the regions undergoing periodic flares can
be isolated. It was found that the flaring methanol and OH masers are, in projection,
about 1600 AU apart. The decay of the flaring OH masers can described in terms
of the variation of the free-free emission from a recombining hydrogen plasma, similar to
what is found for the methanol masers in G9.62+0.20E.

\section{Observations and data reduction}

\subsection{KAT-7 }

The seven-dish Karoo Array Telescope \citep{Foley2016} was built as an engineering
prototype for the 64-dish MeerKAT array in the
Karoo region of the Northen Cape, South Africa.  It consists of 7 12-m diameter dishes
with prime focus linearly-polarised receivers covering a frequency range of 1.2 to 1.95
GHz.  It is a compact array, with a maximum baseline of 186m and shortest baseline of 26m.
The system temperature of the antennas is approximately 30 K, and apperture efficiency is
on average 65\%.

The OH maser monitoring observations started during the early stages of spectral line
commissioning on KAT-7 in February 2013.  We used the c16n2M4k correlator mode, which
gives a velocity resolution of 68 m s$^{-1}$ at the OH rest frequencies.  Since May 2013,
observations at 1665 and 1667 MHz were interleaved in a single schedule block using an
LST-based scheduling mechanism, which ensures consistent uv-coverage from one observation
to the next and ensures that observations at both frequencies are executed
quasi-simultaneously.  Prior to this observations at each frequency were scheduled
separately, preferably within the same day. The total integration time on source at each
frequency was initially 20 minutes and subsequently increased to up to 50 minutes for the
weekly monitoring programme and up to 80 minutes for the daily observations whenever
possible. Daily observations during predicted flares were dynamically scheduled subject to
telescope availability.  Not all antennas were always available, and integration time was
increased when possible to compensate. The typical rms noise achieved ranges from 0.15 to
0.2 Jy. The typical beam size is $\sim$ 3 arcmin thus the masers are unresolved and the
relative positions of the masers cannot be measured due to the poor resolution.

PKS 1934-638 -- the flux and bandpass calibrator -- was observed for five minutes every
hour.  PKS1730-130 was used as the gain calibrator and was observed once every ten minutes
for one minute.

The data were calibrated following standard interferometric calibration procedures using
CASA \citep{McMullin2007}. The bandpass response turned out to be stable in time so all
scans were combined to produce a single bandpass calibration table.  In order to avoid
introducing noise from the bandpass calibration into the spectrum (the continuum emission
is similar in strength to the bandpass calibrator) the bandpass solution was smoothed
using a 3rd-order polynomial in both phase and amplitude.  We were unable to obtain a
reliable polarisation calibration in the narrow band mode due to insufficient signal to
noise on the polarisation calibrator 3C286 so only the Stokes I product is considered
here.

KAT-7 has a 1 square degree field of view and excellent sensitivity to low surface
brightness emission. This results in a complex continuum image. Spectral line data cubes
were imaged after continuum subtraction in the u-v plane. The sheer volume of the data
necessitated automated data reduction and imaging.  The first 'quick-look' image cubes
produced had imaging artefacts (spurious source detections in isolated channels) and
higher rms noise than expected in some channels.  To address this problem, a deep image
was made of the field by combining all available observations at the time - some 60 hours
of data on-source - in order to optimise the imaging parameters.  There are five spectral
line sources in the field of view, including a deep absorption feature, which were fully
characterised to determine the appropriate velocity ranges for continuum subtraction, and
to create a velocity-dependent CLEAN mask for non-interactive image deconvolution. The
clean threshold was determined by measuring the rms noise in an emission-free
channel. This approach led to much better and more consistent imaging quality even when
uv-coverage was reduced.

Data quality was severely compromised by solar interference during December and early
January when first the gain calibrator and then the target were less than 12 degrees from
the Sun.  We have discarded all data taken during this period.

\subsection{Single dish observations}
Concurrent monitoring of the methanol and water masers was done using the Hartebeesthoek
Radio Astronomy HartRAO 26-m telescope, subject to scheduling constraints.

Flux calibration at 6.7 and 12.2 GHz was done by daily drift scans on Virgo A and Hydra A
using the flux scale of \citet{Ott1994}.  Amplitude corrections due to pointing errors
were calculated by offset observations to the east-west and north-south half-power points.
Bandpass calibration at 6.7 and 12.2 GHz was done by frequency switching, while
position-switching was done at 22 GHz to accommodate a potentially wider velocity spread.
The pointing offsets at 22 GHz were greater than half a beamwidth at times, making it
impossible to derive reliable amplitude correction since we detect the source in only one
or two of the offset pointing positions.  These observations have been discarded from the
timeseries.  The 22 GHz flux densities were also corrected for atmospheric absorption
using water vapour radiometer and environmental data. Jupiter was used as a flux
calibrator.

\section{Results}
\begin{figure}
	\includegraphics[width=\columnwidth]{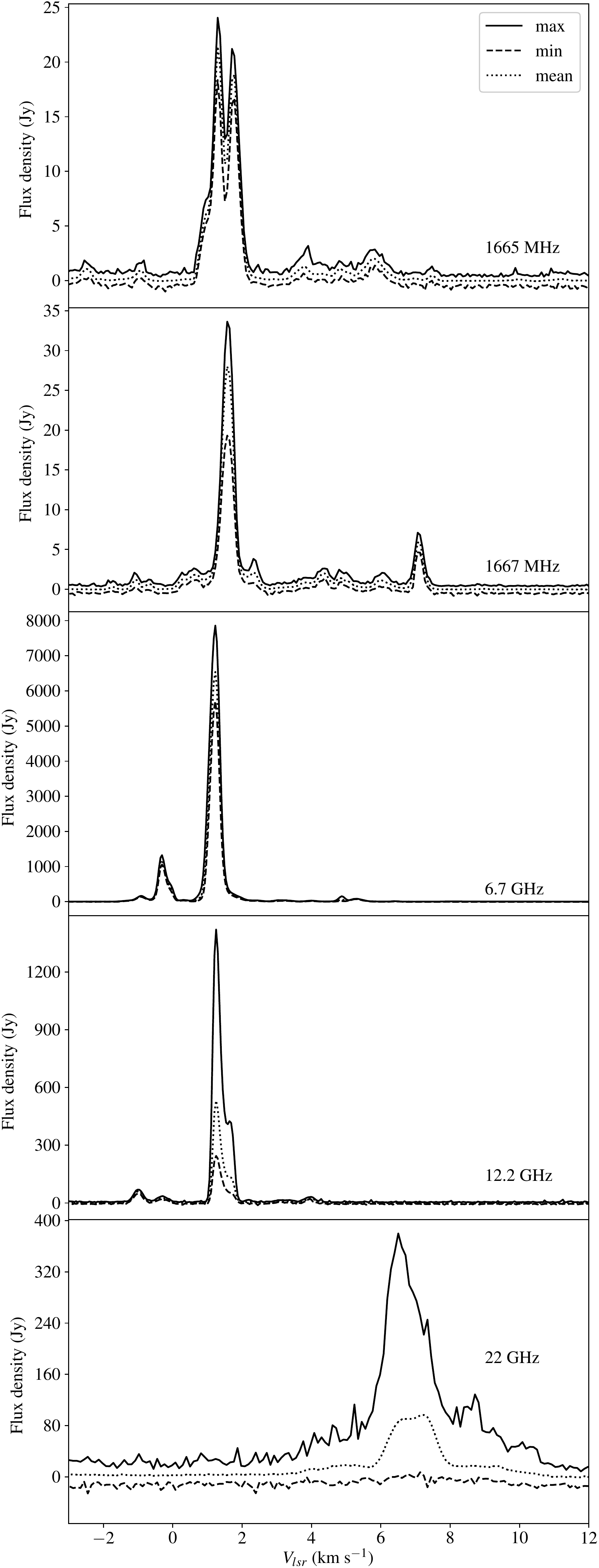}
    \caption{Spectra of the maser transitions monitored, showing the range of variation in
      each velocity channel.}
    \label{fig:spectra}
\end{figure}

Figure~\ref{fig:spectra} shows the spectra and range of variation of the OH masers at 1665
and 1667 MHz, the methanol masers at 6.7 and 12.2 GHz and the water masers at 22 GHz.
Variability is seen in the same velocity ranges for the methanol and hydroxyl transitions,
between 1 to 3 \kms.  The water masers are highly variable but have a different and
larger velocity range.

\subsection{Hydroxyl maser time-series}

\begin{figure}
	\includegraphics[width=\columnwidth]{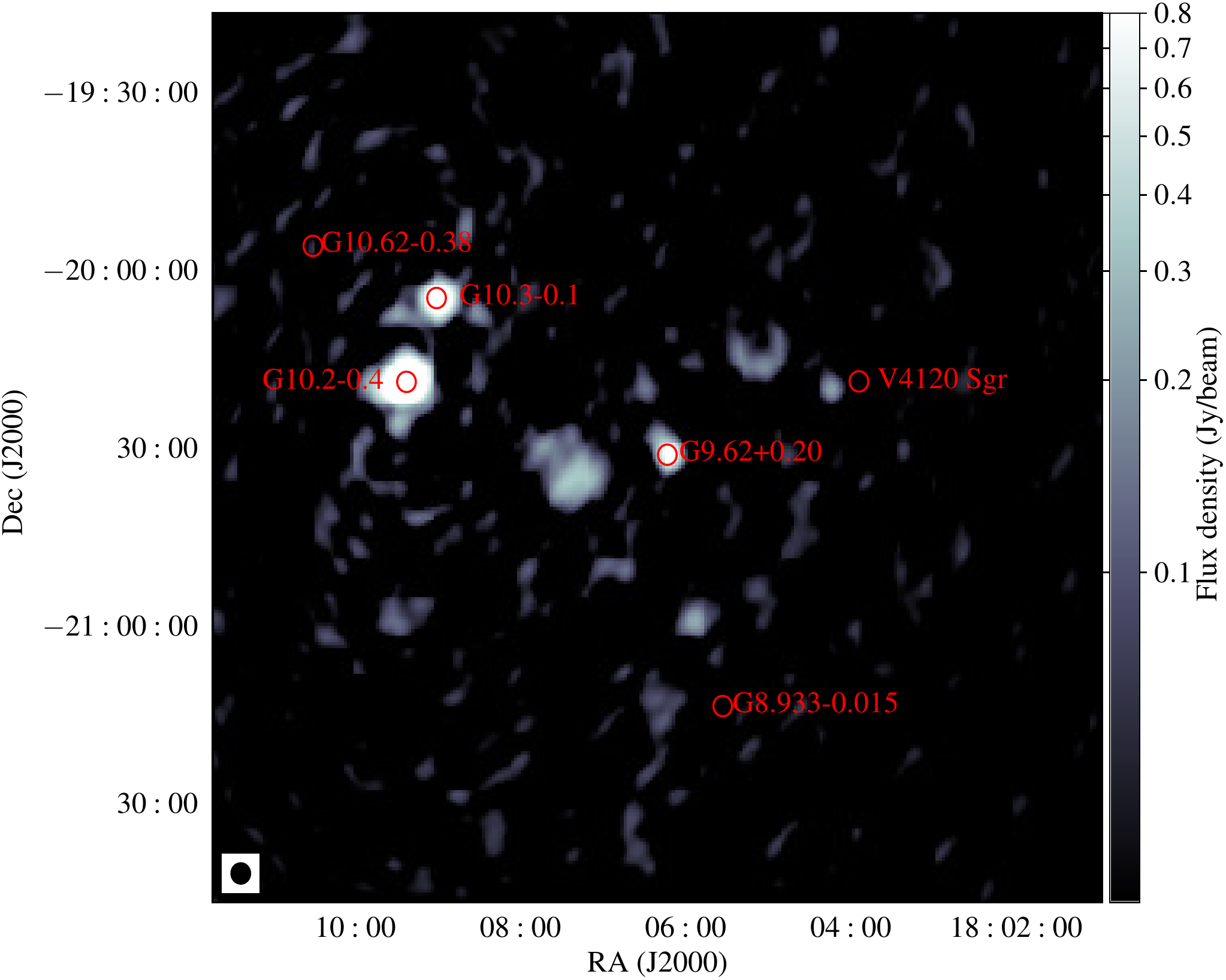}
    \caption{Overview of sources detected in the field.  The colour-scale shows the
      continuum image, Locations of spectral line sources are indicated by the red
      circles. The size of the circles indicates the synthesised beam size.}
    \label{fig:cont}
\end{figure}

Figure~\ref{fig:cont} shows the continuum image of the field with locations and
identification of spectral line sources marked in red.  Note that primary beam correction
has not been applied to this image. At L-band we are detecting optically thin, evolved
\ion{H}{II} regions.  Continuum emission is visible at the G9.62+0.20 site and is most
likely dominated by components A and B \citep{Garay1993}, which are more evolved and 
extended \ion{H}{II} regions in the complex.

\begin{figure*}
	\resizebox{0.8\textwidth}{!}{\includegraphics{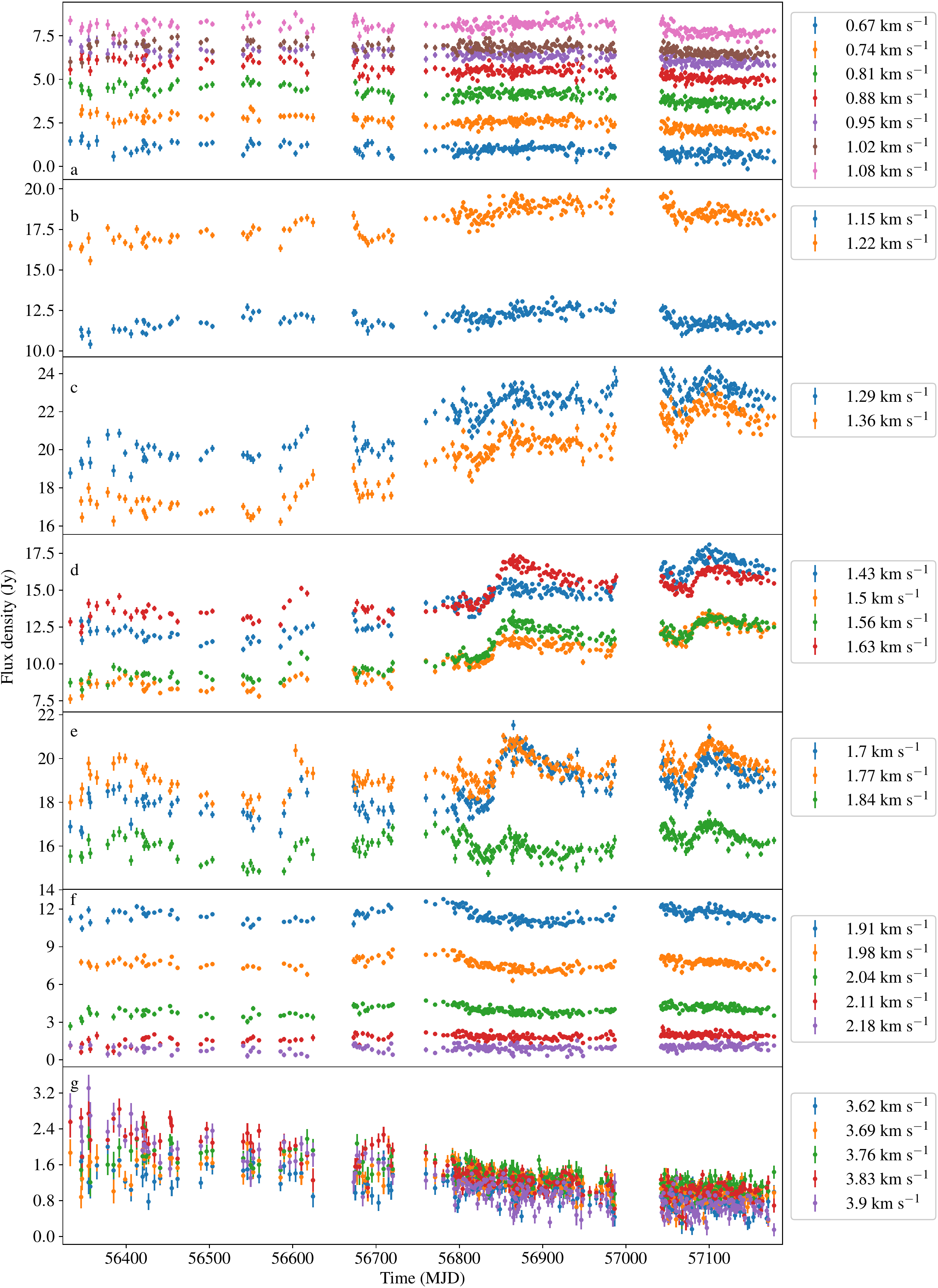}}
    \caption{Time-series of individual velocity channels for the 1665 MHz OH masers.
      Plots have been split across channels to aid in visual identification of correlated
      behaviour.}
    \label{fig:1665_ts}
\end{figure*}

\begin{figure*}
	\resizebox{0.8\textwidth}{!}{\includegraphics{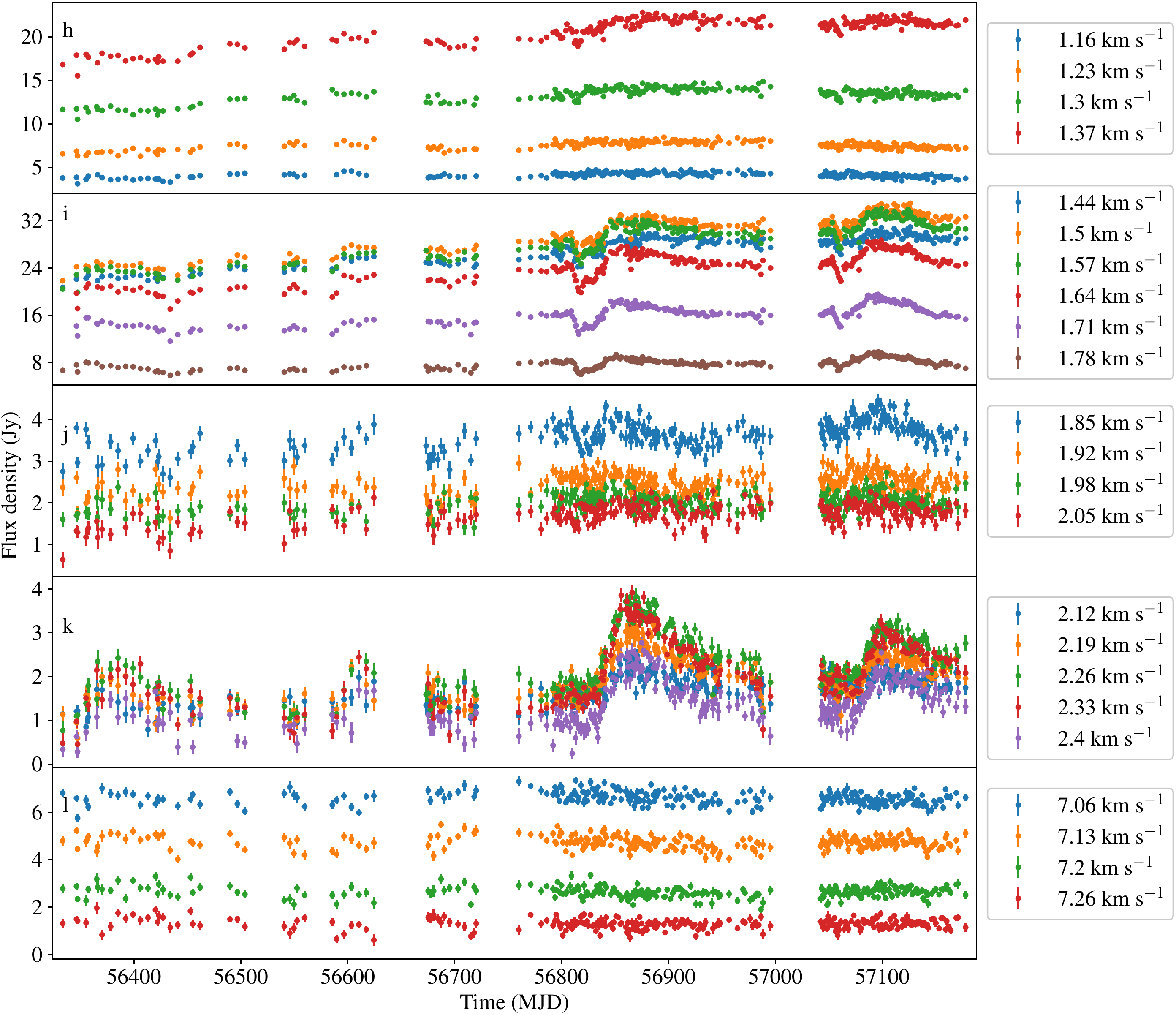}}
    \caption{Time-series of individual velocity channels for the 1667 MHz OH masers.
      Plots have been split across channels to aid in visual identification of correlated
      behaviour.}
    \label{fig:1667_ts}
\end{figure*}

Time-series were generated by fitting a 2-dimensional Gaussian to channels of interest in
the image plane.  The uncertainty in the measured amplitude was calculated by measuring
the rms noise in an outer quadrant of the image for that channel.  We generated
time-series for all velocity channels which had amplitudes greater than three times the
rms noise.  Figures~\ref{fig:1665_ts} and \ref{fig:1667_ts} show the time-series for the
1665 and 1667 MHz data, respectively.  The KAT-7 data suffers from heavy spectral
blending, which could not be resolved by fitting individual Gaussian components to the
spectra.  Instead we plot the measured flux density in each channel.  We have grouped
channels by visual inspection of both the spectral structure and the characteristics of
the variability, which appears markedly different as we move between spots in the general
star forming region.  The most notable behaviour is seen in the 1667 MHz transition in the
peak velocity channels at $\sim$ 1.7 \kms. These features show a drop in intensity at the
time that the methanol masers start to flare, followed by a sharp rise, and a gradual
decay. In contrast, the features at $\sim$ 2 \kms~ show a flare profile that is more
similar to that of the methanol masers.

\subsection{Comparison with VLBI spectra and radio continuum positions}

\begin{figure*}
	\resizebox{\hsize}{!}{\includegraphics{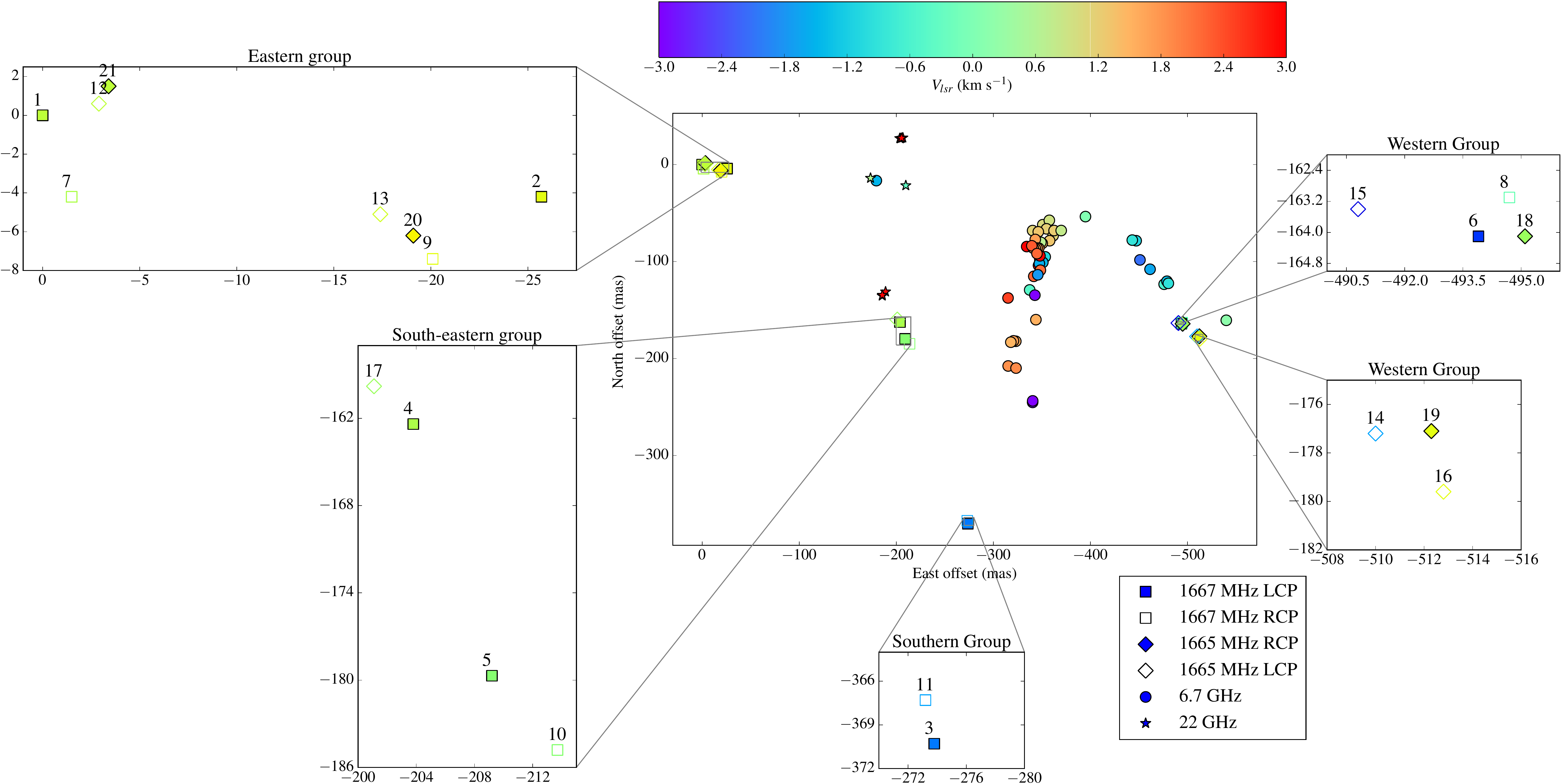}}
    \caption{Maser spot positions as reported in Table 2 of \citet{Sanna2015}.  }
    \label{fig:spotmap}
\end{figure*}

Since the individual hydroxyl maser spots cannot be resolved by KAT-7 and there are
multiple spots with overlapping velocities, we inspected the VLBI spectra more closely to
see if they could be correlated with the temporal and velocity behaviour seen in the KAT-7
data. The radio continuum source E appears to have two components - E1 is the stronger
peak over which the methanol masers are centred, while E2 is one-fifth of the strength of
E1 and is located $\sim$ 1000 AU in projection to the north-east
\citep[see][]{Sanna2015}.

Figure~\ref{fig:spotmap} shows the maser spotmap from \citet{Sanna2015}, with zoomed
insets on the various OH maser clumps.  The OH masers are not co-located with the methanol
masers; instead they are distributed in several clumps around the region. The clump to the
north-east is on the far side of component E2. Other clumps are distributed offset from
the peak of E2 to the south-east, south and west.  The water masers are located between E2
and E1. It is also known that not all of the methanol masers flare \citep{Goedhart2005},
so comparison of the maser positions and their time-dependent behaviour may help to narrow
down the origin of the periodic behaviour.

\begin{figure}
	\includegraphics[width=\columnwidth]{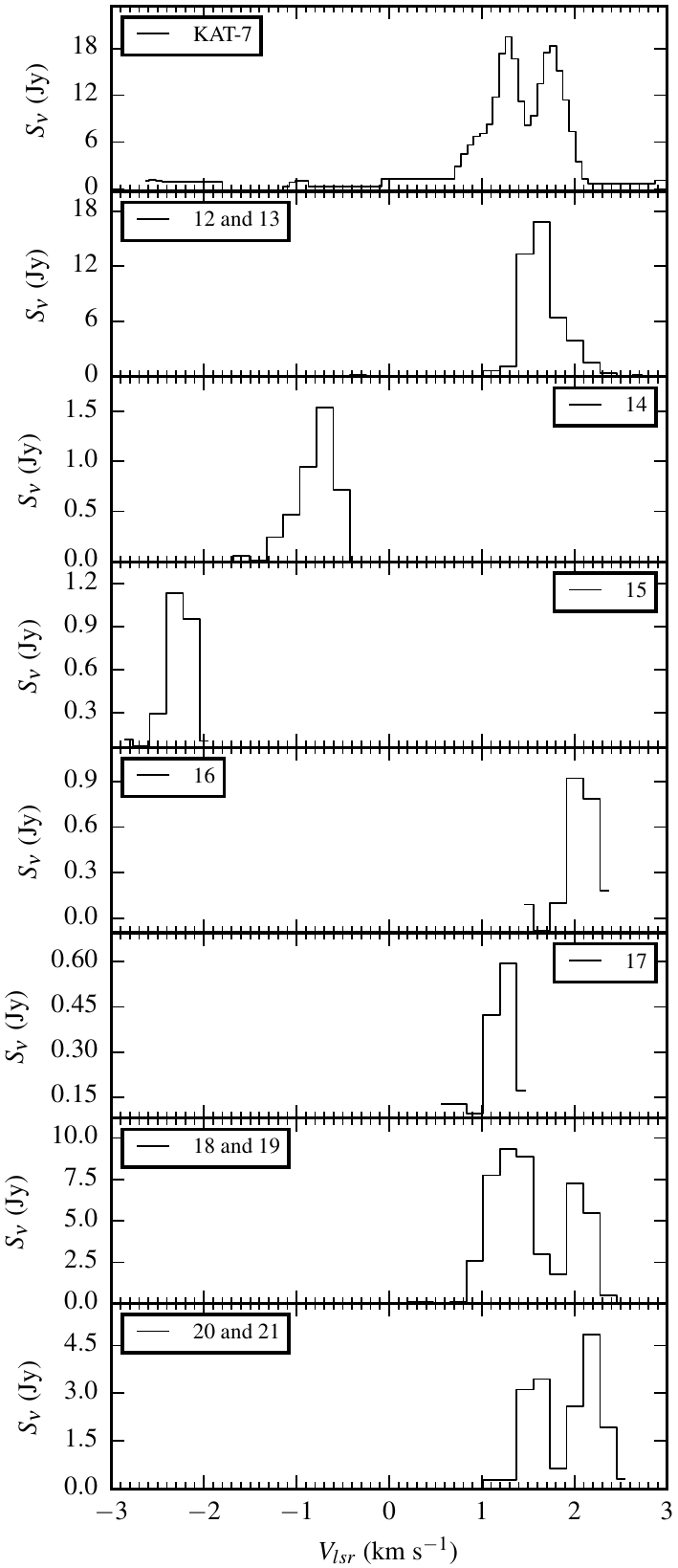}
    \caption{Spectra of individual 1665 MHz maser features from \citet{Sanna2015}. The
      top panel shows the KAT-7 spectrum from the nearest epoch (2013-07-16) 14 days
      later.}
    \label{fig:vlbi_spec1665}
\end{figure}

\begin{figure}
	\includegraphics[width=\columnwidth]{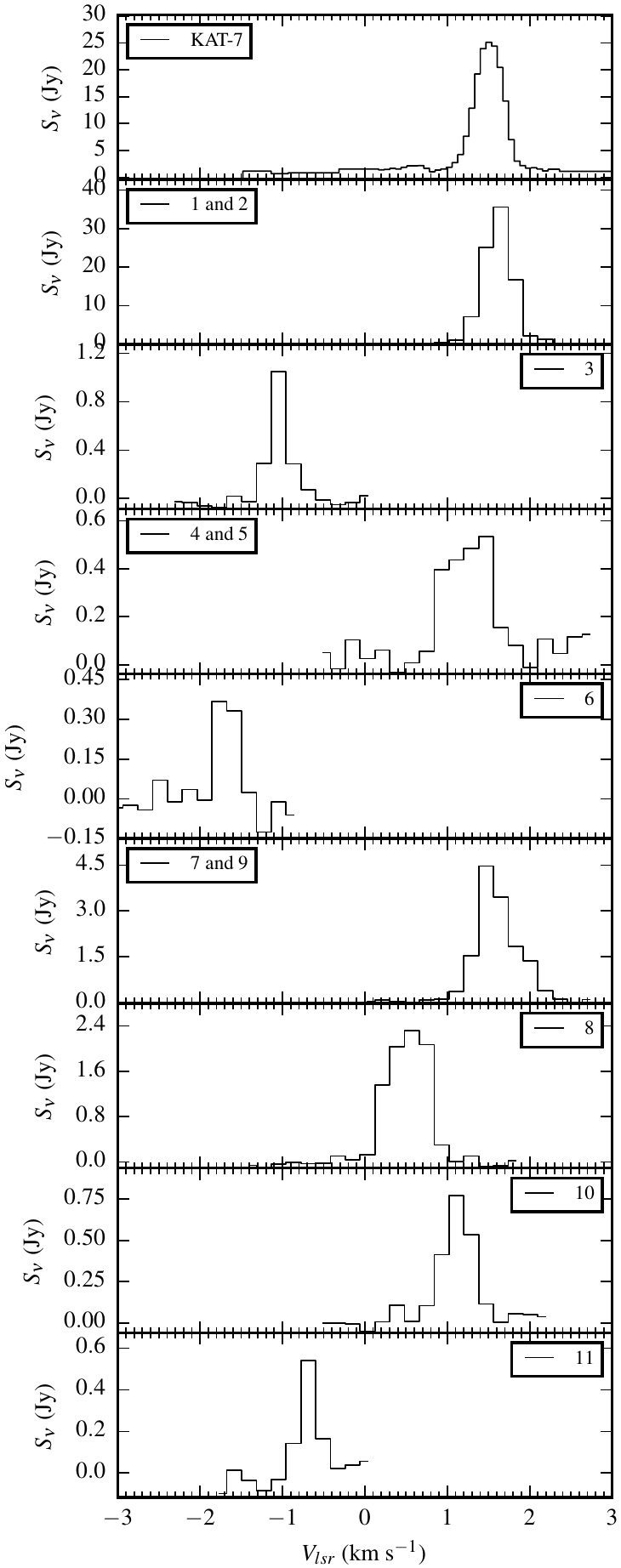}
    \caption{Spectra of individual 1667 MHz maser features from \citet{Sanna2015}. The
      top panel shows the KAT-7 spectrum from the nearest epoch (2013-07-16) 14 days
      later.}
   \label{fig:vlbi_spec1667}
\end{figure}

Figures~\ref{fig:vlbi_spec1665} and \ref{fig:vlbi_spec1667} show the VLBA spectra, with
the KAT-7 spectra taken at the closest epoch (16 July 2013) for comparison.  These spectra
are used to inform the allocation of spectral channels to mapped features.  The following
discussion will work through the enumerated panels in Figures~\ref{fig:1665_ts} and
\ref{fig:1667_ts}. Panel `a' covers velocities in the range 0.67 to 1.08 \kms (the
shoulder on the 1665 MHz spectrum) and shows relatively constant flux densities throughout
the monitoring period.  These channels may be part of features 18/19 in the western group
but the VLBI spectra show a single channel of 2.5 Jy in this velocity range while the the
KAT-7 spectra show far more flux. It is not completely clear where panel `b' channels
originate. It may be a blend of spatially separate features.  The velocity range covered
by panels `b' and `c' -- 1.15 to 1.36 \kms -- is the same as that of features 17, 18 and
19 in the western group, however the VLBI spectra recover significantly lower flux. It may
be that the weak indication of flaring behaviour in panel `c' is spectral blending with
features 12 and 13 shown in panel `d', and that the western group does not flare.  The
features in panel \q{e} are unambiguously in the eastern group.  Panel `e' probably shows
the blending of features 12/13 with features 18 and 19 in the west. In panel `f', note the
slight rise in flux density at 1.9 to 2.04 \kms during May-June 2014 and a possible
similar event February 2015, prior to the expected periodic flare.  These features are in
the same velocity range as 16, 18, 19 to the west, or 20 and 21 in the east. The feature
in panel `g', 3.62 to 3.9 \kms, appears to arise from a region $\approx$ 3 arcsec to the
south of the peak of the \ion{H}{II} region \citep{Fish2005} and are not shown in
\citet{Sanna2015}. These features do not show any sort of correlated variability with the
other masers.

Now to consider the 1667 MHz transition. Here the spectrum is simpler and the variability
much more pronounced. Panel `h' does not show any significant variation.  It is not
entirely clear where this emission arises since the VLBI spectra at this velocity range,
1.16 to 1.38 \kms, do not recover much flux.  It could belong to features 4, 5 or 10, all
of which are in the south-eastern group. Panel `i' in the range 1.44 to 1.78 \kms, shows
very well demarcated drops in intensity, which occur at the onset of the methanol flare,
as we will show in the next section.  After the `dip', the masers increase in strength to
a level higher than the pre-flare level then slowly recover over a period of several
months.  Features in this velocity range include 1, 2, 4, 5, 7, 9 and 10.  However,
considering peak flux densities the contribution must be predominantly from features 1, 2,
7 and 9 which are in the east. Panel `j' does not show much variation. This velocity
range, 1.85 to 2.05 \kms, may be covered by features 7 and 9 but its not clear why they
would not flare if all the other masers in this region are flaring. Panel `k' shows very
pronounced flaring behaviour in the velocity range 2.12 to 2.4 \kms~ but it is unclear
where these features are in the VLBI map - they could be part of 7 \& 9 or 4 \& 5 in the
south-east.  The features in panel `l', as with panel `g', arise from a region 3 arcsec to
the south \citep{Fish2005}.

To summarise the most variable features appear to be in the eastern group, closer to \ion{H}{II}
region component E2 and it is likely that none of the other OH maser groups are flaring.
 
\begin{figure}
	\includegraphics[width=\columnwidth]{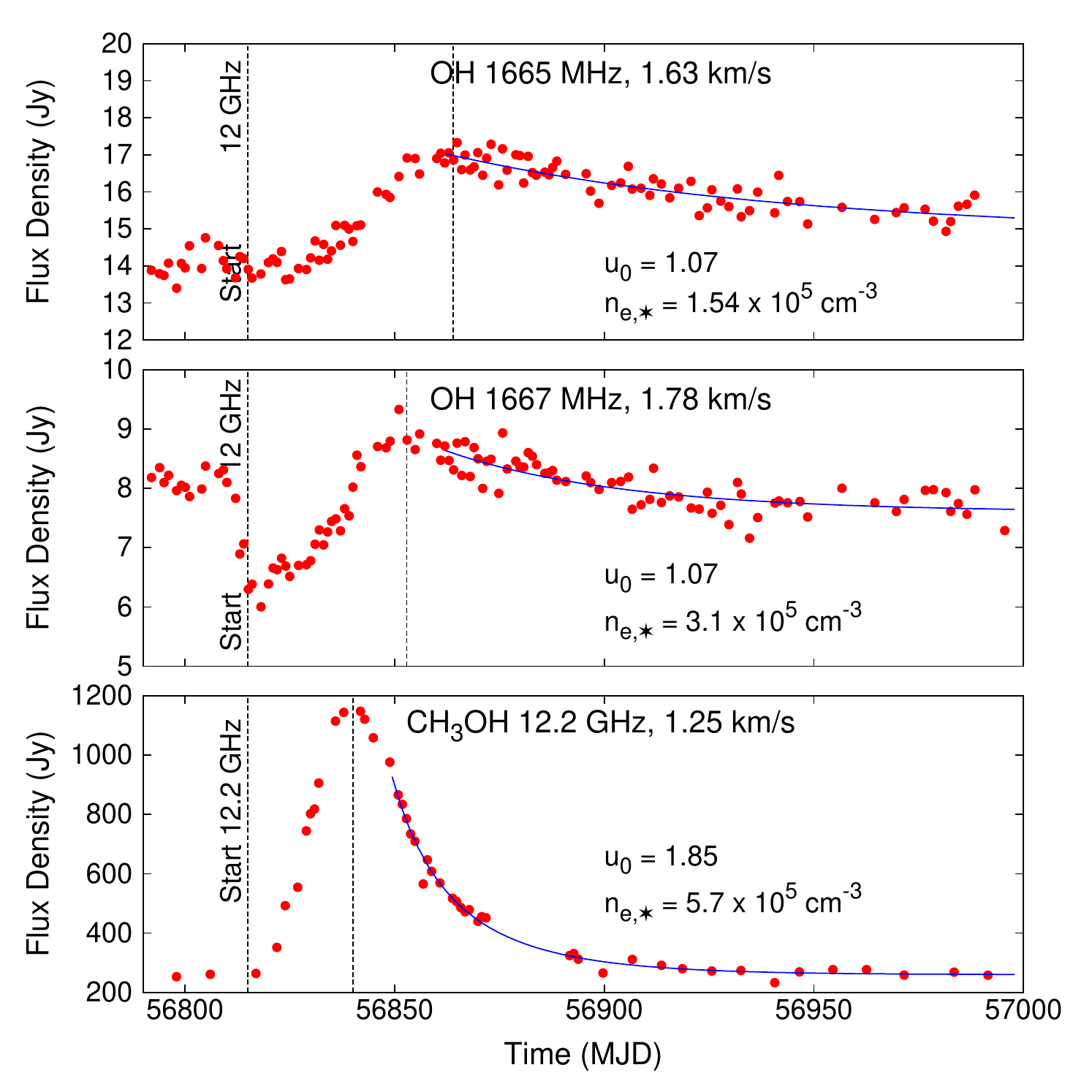}
    \caption{More detailed comparison of the profiles of the OH and methanol maser flares.
      The solid blue lines are the fits of eq. A7 of \citet{VanderWalt2009} to the decay
      of the flares. See \citet{VanderWalt2009} for the interpretation of $u_0$ and
      $n_{e,\star}$.}
   \label{fig:meth_comp2}
\end{figure}

\subsection{Comparison of OH and methanol maser flare profiles}

As already noted above, the OH maser flares have a ``characteristic'' dip in the flux
density before the flare. Such behaviour has not been reported for the methanol maser
flares. In Figure~\ref{fig:meth_comp2} a more detailed comparison between the OH and
methanol maser flares is shown. The dip is seen to be significantly more pronounced for
the 1667 MHz maser than for the 1665 MHz maser. It is rather interesting to note that for
the 1667 MHz maser the onset of the dip is very near to the start of the 12.2 GHz maser
flare. After the dip, the OH masers increase to a maximum after which they decay to the same
level as before the onset of the 12.2 GHz maser flare. The time intervals between the peak
of the 12.2 GHz maser and when the 1665 and 1667 MHz masers reach their maxima are about
23 and 13 days respectively.

Given that after reaching a maximum, the OH masers recover to the same pre-flare level, a
behaviour also seen for the methanol masers, we also tried to fit the decay part of the OH
masers with equation A7 of \citet{VanderWalt2009} as well as to the decay of the 12.2 GHz
methanol maser flare shown in Figure~\ref{fig:meth_comp2}. The fits are shown as the blue
solid lines in Figure~\ref{fig:meth_comp2}. It is seen that for all three masers the decay
of the masers are described very well by eq. A7 of \citet{VanderWalt2009}. Assuming that
the decay of the masers are indeed related to the recombination of a partially ionized
hydrogen plasma, the fit allows us to obtain estimates of the quiescent state electron
density ($n_{e,\star}$) as well as the ratio ($u_0$) of the electron density from where
the recombination started and $n_{e,\star}$. The values of these two quantities obtained
from the fits are shown in the respective panels of Figure~\ref{fig:meth_comp2}. The
derived quiescent state densities range from $5.7 \times 10^5~ \mathrm{cm^{-3}}$ for the
12.2 GHz maser at 1.25 km\,s$^{-1}$ to $1.54 \times 10^5~\mathrm{cm^{-3}}$ for the 1665
MHz OH maser. The order of magnitude of these densities are what is expected for very
young \ion{H}{II} regions. It is also seen that for the OH masers the value of $u_0$ as
found from the fit is only 1.07, in agreement with the small amplitude of the flares. For
comparison, $u_0 = 1.85$ for the strong 12.2 GHz flare.

\subsection{Water maser monitoring}

\begin{figure*}
	\resizebox{\hsize}{!}{\includegraphics{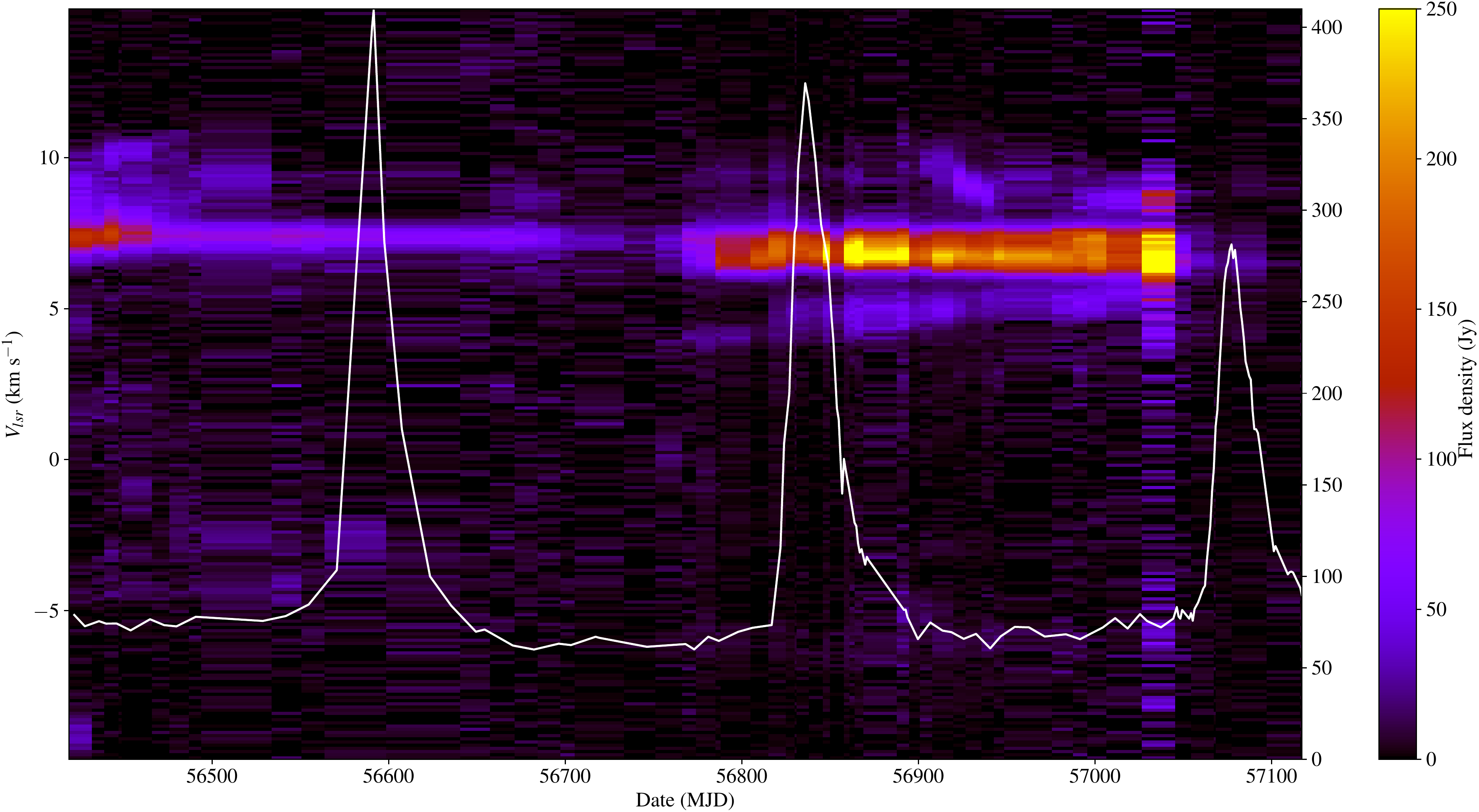}}
    \caption{Flux density of water masers as a function of time and velocity in the
      colorscale plot. The white curve is the 12.2 GHz methanol for comparison.  }
    \label{fig:h2o}
\end{figure*}

Figure~\ref{fig:h2o} shows a dynamic spectrum of the water maser observations.  While
the masers are variable, there does not appear to be any particular correlation with the
methanol or hydroxyl, but the time-coverage during the September 2013 flare is extremely
sparse due to bad telescope pointing.  There are some transient features at $\sim$ 5 \kms~
and $\sim$ 8 \kms~ in the second flare cycle, which also appear to have slight changes in
peak velocities very typical of water masers in an outflow.

\begin{figure}
	\includegraphics[width=\columnwidth]{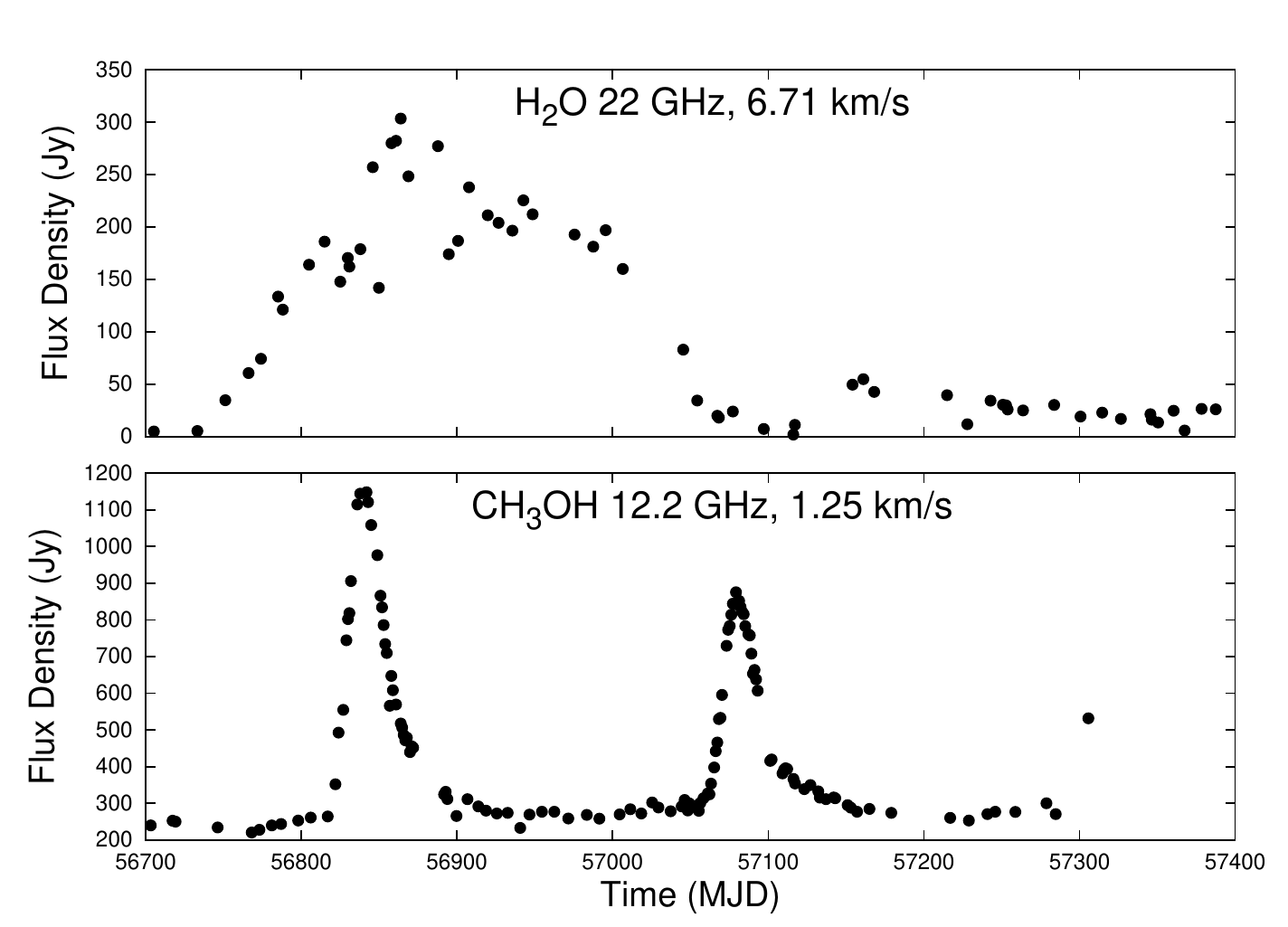}
    \caption{Comparison of water and 12.2 GHz methanol maser flares}
   \label{fig:methwater}
\end{figure}

In Figure~\ref{fig:methwater} we show a more detailed comparison between the time series
of the water- and 12.2 GHz methanol masers between MJD 56700 and 57400.  It is seen that
the duration of the water maser flare covers a time interval approximately equal to the
period of the 12.2 GHz flares. The water and methanol maser flares do not seem to be
related in the same way as seems to be the case for the OH and methanol maser flares.

\section{Discussion}

There are quite a number of interesting aspects to note in G9.62+0.20.  We first note, using Figure 1 of \citet{Sanna2015}, that, in projection, the periodic
methanol and OH masers are located approximately 330 and 1930 AU from the continuum
peak of the VLA A-array 7mm emission. Although there are OH masers located closer to the
continuum peak, these do not appear to be periodic at all. The question then is why the
periodic OH masers are located so much further than the periodic methanol masers from the
peak of continuum. As shown above, equation A7 of \citet{VanderWalt2009} provides a very
good fit to the decay of the 12.2 GHz methanol maser flare. The implication is that the
variation in the electron density occurs at such a position that the \ion{H}{II} region is
optically thin for outward propagating 12.2 GHz photons in the direction of the masing
gas. Since the optical depth is proportional to $\lambda^{2.1}$, it follows that at the
projected position of the periodic methanol masers, the optical depth at 1665 MHz is about
65 times greater than at 12.2 GHz. If, for example, the optical depth at 12.2 GHz is
0.1, it means that it is 6.5 at 1665 MHz and therefore that all variability will be damped
due to absorption in the \ion{H}{II} region. Qualitatively, it is therefore expected that
if there are periodic OH masers, that these will be located further from the core of the
\ion{H}{II} region compared to the periodic methanol masers. This might explain the
relative projected positions of the periodic methanol and OH masers in G9.62+0.20E.

Perhaps the most intruiging aspect of the periodic OH masers in G9.62+0.20E is the
pronounced sharp decrease (dip) in the maser flux density before the ``flare''. That the
dip is a real feature is clear from the fact that it is seen in two flares and is also
present in the 1665 MHz masers. As noted above, the periodic methanol masers do not show
this behaviour. If, as proposed by \citet{VanderWalt2009}, \citet{VanderWalt2011} and,
\citet{vanderwalt2016}, the periodicity of the methanol masers in G9.62+0.20E are driven
by a colliding-wind binary system and that the flares are due to changes in the background
free-free emission from the \ion{H}{II} region, it is required that the dip should also be
explained within the same framework. We consider two possibilities.

(a) \textit{The periodicity is due to a CWB associated with component E1:} Although no
quantitative explanation can be given, qualitatively we note the following.  First, from
the maps of \citet{Sanna2015} it is seen that the periodic OH masers are projected
significantly ($\sim$1600 AU) further from the core of the \ion{H}{II} region E1 than the
periodic methanol masers. Should both masers amplify the free-free emission from the
background \ion{H}{II} region, it follows that they probe two completely different parts
of the \ion{H}{II} region. The ionization structure of an UC\ion{H}{II} region as
calculated with the photo-ionization code {\tt Cloudy} \citep[see eg.][]{VanderWalt2011}
shows that partially ionized gas extends beyond the ionization front. Within the framework
of the CWB scenario, the lower energy ionizing photons from the hot shocked gas are
absorbed at the ionization front and gives rise to the flaring of the methanol
masers. Higher energy ($>$ 100 eV) photons, on the other hand, can propagate beyond the
ionization front due to the significantly lower photo-ionization cross section for
HI. Ionization of hydrogen by these photons will result in higher average electron
temperatures. In the optically thin case the free-free emission is proportional to the
volume emissivity which, in turn, is proportional to $T_e^{-0.5}$. Raising the electron
temperature of the plasma thus lowers the volume emissivity.

A \textit{possible} explanation then for the dip in the 1667 MHz maser flux density is that,
because of the higher energies of the ionizing photons, the electron temperature is raised
significantly during the ionization event in the gas against which the OH maser is
projected, resulting in a decrease in the free-free emission. After the pulse of ionizing
photons has passed, the electrons cool to the equilibrium temperature giving rise to an
increase in the free-free emission and therefore also in the maser emission. Simultaneous
to the increase in the electron temperature, the electron density also increases due to
the pulse of ionizing photons. The exact profile of the flare (which includes the dip)
depends on the magnitude of the change in electron temperature, the cooling rate of the
electrons and the ionization rate associated with the pulse of ionizing photons. The
observed peak of the OH maser flares therefore does not necessarily indicate the time of
the peak of the pulse of ionizing photons as in the case of the methanol maser flares. The
difference in time between the peaks of the methanol and OH maser flares, might therefore
not be due to a geometric delay only (see below) but also to other physical effects. The
required change in electron temperature to explain the magnitude of the dip can be
estimated by considering the ratio of flux densities of the 1667 MHz maser before the dip
and at the minimum of the dip (Figure~\ref{fig:meth_comp2}).  It then follows that the
electron temperature has to be raised by a factor of 1.8 to explain the observed decrease
in the maser flux density.  Thus, for example, if before the dip the electron temperature
was 5000 K, it has to be raised to 9000 K in order to explain the decrease seen in the
maser flux density.

(b) \textit{The periodicity is associated with component E2.} Given that the decay of all
three masers are described very well in terms of the recombination of a hydrogen plasma,
it is reasonable to assume that component E2 must then also be a periodic source of
ionizing photons which influence the flux of seed photons for both the methanol and OH
masers. Since the methanol masers are projected against E1, it is therefore required that
the ionizing photons propagate from E2 to the position against which the periodic methanol
masers are projected. However, it is clear from \citet{Sanna2015} that there is very
little ionized gas associated with E2, which means that a significant fraction of the flux
of the periodically produced ionizing photons will be absorbed close to E2. It is therefore
difficult to see how a source of periodically varying ionizing photons located at E2 will
influence that part of E1 against which the periodic methanol masers are projected. 

\begin{figure}
	\includegraphics[width=\columnwidth]{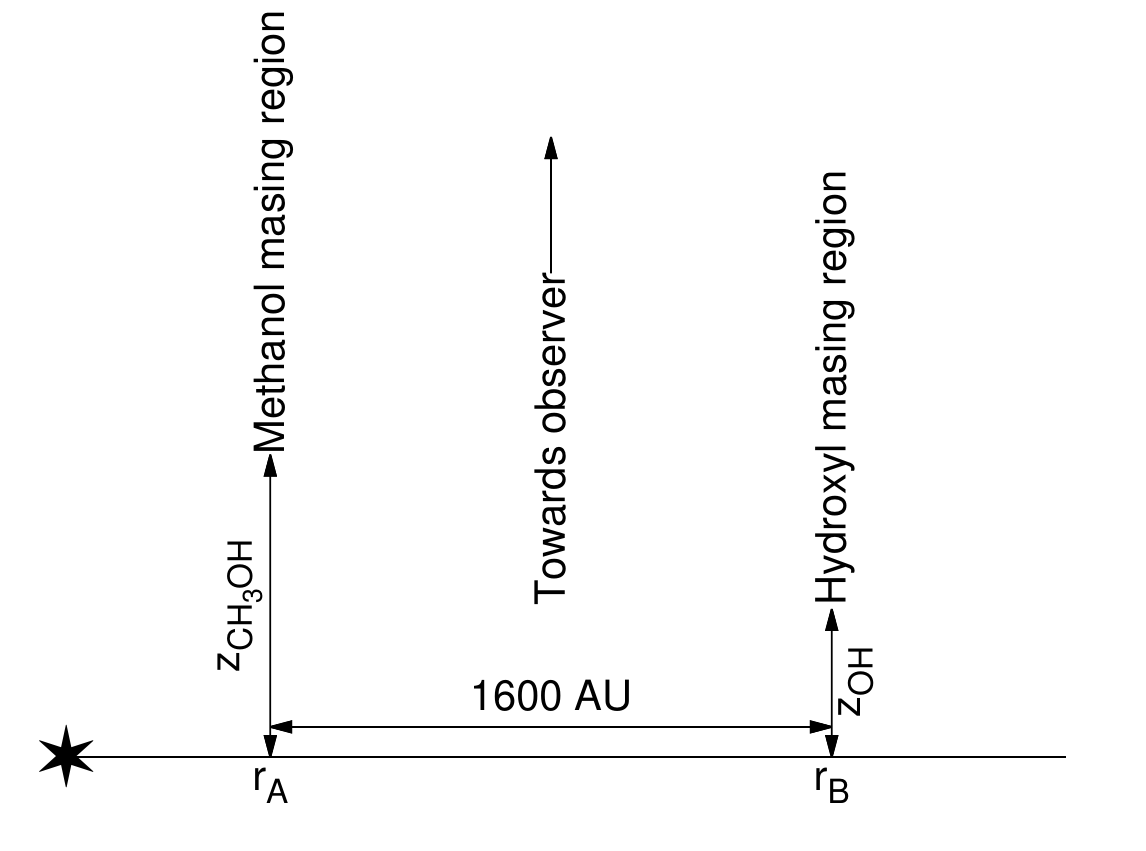}
    \caption{Simplified geometry for the locations of the periodic methanol and hydroxyl
      masers. The ionizing star is indicated by the star symbol. The ``surface'' of the
      \ion{H}{II} region is assumed to be flat. The methanol and hydroxyl masers
      respectively respond to changes in the free-free emission from points A and B}
   \label{fig:geometry}
\end{figure}

Having excluded E2 as the driving source for the periodic methanol and OH masers, we
finally note the following. From Figure~\ref{fig:meth_comp2} it is seen that although the
peak of the 1667 MHz flare lags that of the 12.2 GHz flare by about 13 days, the dip of
the 1667 MHz maser occurs almost simultaneously with the onset of the 12.2 GHz methanol
maser flare. In view of the above possible explanation for the dip, it seems reasonable to
regard the onset of the 12.2 GHz flare and the almost simultaneous sharp decrease of the
1667 MHz maser flux density to be caused by the same ionization event. However, due to the
difference in projected distances from the core of the \ion{H}{II} region, we then expect
the dip of the OH maser flares to be delayed by about nine days with respect to the onset
of the 12.2 GHz flare. Considering the very simple geometry in Figure \ref{fig:geometry},
the time difference between the methanol and hydroxyl flares is given by $\Delta t =
(r_{\mathrm{B}} - r_{\mathrm{A}} + z_{\mathrm{OH}} - z_{\mathrm{CH_3OH}})/c$. For the
present case $\Delta t \sim 0$, from which follows that $z_{\mathrm{CH_3OH}}-
z_{\mathrm{OH}} \approx r_{\mathrm{B}}- r_{\mathrm{A}}$. Our observations of the methanol
and OH maser flares therefore suggest that the masing region of the periodic 12.2 GHz
methanol masers are located about 1600 AU further from the ``surface'' of the \ion{H}{II}
region into the molecular envelope compared to the location of the periodic OH masers.

As noted above, a water maser flare occured that overlapped in time with a 12.2 GHz
methanol maser flare. It seems as if a comparison between the water and methanol maser
flares similar to that between OH and methanol masers cannot be made. Although the water
maser flare overlaps in time with the 12.2 GHz methanol maser, the water maser flare
clearly starts before the methanol maser flare. Also, the decay time of the water maser
flare is significantly different from that of the OH and methanol maser flares and does not suggest
any causal relation with the methanol and OH flares.  

To our knowledge, the results presented above are the first conclusive observational evidence of periodic variability of OH masers associated with a high-mass star forming region. \citet{Green2012} found an indication of periodicity in the OH masers toward G12.889+0.489 (which has a 29.5 day period in methanol), but the time-series were too undersampled to obtain detailed cycle profiles and no contemporaneous methanol monitoring was done. It is interesting that the hydroxyl masers appear to  undergo a drop in emission coincident with the expected minima of 6.7 GHz methanol masers but a key difference is that the hydroxyl masers do not show any flares. G12.889+0.489 also differs from G9.62+0.20E in that it has somewhat irregular flaring behaviour in the methanol, but shows a well-defined minimum, which seems to be periodic \citep{Goedhart2009}, while the peaks of the methanol flares can occur any time in an 11-day window. It is quite likely, given the short period and the difference in the methanol maser light curves, that a different mechanism is modulating the maser intensity. It would undoubtedly be of great benefit to our understanding of these phenomena to conduct intensive monitoring of both maser species through one 29.5 day cycle in this source.

\section{Summary and Conclusions}

We presented the first conclusive observational evidence for the periodic variability of OH mainline
masers associated with a high-mass star forming region. The 1667 MHz masers show a
pronounced dip in flux density which closely coincide with the onset of the 12.2 GHz
methanol maser flare. The decay of the OH maser flares, similar to that of the methanol
maser flares, can be described very well by the decrease in the free-free emission from
the background \ion{H}{II} region as expected from a recombining hydrogen plasma. A
possible explanation, within the framework of the colliding-wind binary scenario, for the
dip in flux density of the 1667 MHz masers is that it is due to an increase in electron
temperature following the ionization of the outer regions of the \ion{H}{II} region by
photons with energy greater than about 100 eV. We also argued, within this framework that
the masing region for the periodic 12.2 GHz methanol masers are located about 1600 AU
further from the ``surface'' of the \ion{H}{II} region into the molecular envelope compared to
the location of the periodic OH masers.

%section*{Acknowledgements}

%%%%%%%%%%%%%%%%%%%%%%%%%%%%%%%%%%%%%%%%%%%%%%%%%%



\bibliographystyle{mnras}
\bibliography{goedhart_oh_mon_rev1_clean}

%%%%%%%%%%%%%%%%%%%%%%%%%%%%%%%%%%%%%%%%%%%%%%%%%%

% Don't change these lines
\bsp	% typesetting comment
\label{lastpage}
\end{document}